# RADIATIVE $^3$He($^2$H,$\gamma$)$^5$Li CAPTURE AT LOW DEUTERIUM ENERGY

Sergey Dubovichenko[1*], Nataliya Burkova[2†], Albert Dzhazairov-Kakhramanov[1‡]

[1] Fesenkov Astrophysical Institute "NCSRT" NSA RK, Almaty
[2] Al-Farabi Kazakh National University, Almaty
*dubovichenko@mail.ru
†natali.burkova@gmail.com
‡,ⁱalbert-j@yandex.ru

**Abstract:** The results are presented on the total cross sections, astrophysical *S*-factor, reaction rate of the deuteron radiative capture on $^3$He at the temperatures from 0.03 up to 3 $T_9$ calculated in the framework of the potential cluster model with the forbidden states coming from the classification of the orbital states by the Young diagrams. Within the used model and exploited Young diagram classification made it possible to reproduce the general features of the available experimental data on the cross section and reconstructed astrophysical *S*-factor in the energy range from 200 keV up to 1.4 MeV. The parametrization of the obtained reaction rate has been found and been compared with some other actual reactions with light clusters.

## 1. Introduction

We see the interest to the radiative capture reactions in the isobar-analogue channels $^3$H($^2$H,$\gamma$)$^5$He and $^3$He($^2$H,$\gamma$)$^5$Li is due to following two main reasons. The new data may be found in [1] on their application in the diagnostics of nuclear fusion efficiencies of $^3$H($^2$H,n)$^4$He and $^3$He($^2$H,p)$^4$He reactions used for study of tokamak plasmas in experiments on JET and ITER. These reactions are also parts of nucleosynthesis chain of the processes occurring on the early stage of stable stars formation, as well as possible candidates for the overcoming of the well-known problem of the *A* = 5 gap in the synthesis of light elements in primordial Universe [2].

Present paper reports on the treating of $^3$He($^2$H,$\gamma$)$^5$Li reaction as less examined from our point of view. Thus, near all, but few theoretical model research of this channel are basing on the data of the only one experimental work by Buss, 1968 y. [3] on the total cross section of the deuteron radiative capture on $^3$He in the deuteron energy range 200–1400 keV. The most complete nuclear data base EXFOR [4], and known in open access data basis on the nuclear characteristics PHYSICS, CDFE, NASA DATA (see, for example, [5,6]) contain also these very data only. There are also no consistent detail study of $^3$He($^2$H,$\gamma$)$^5$Li reaction including the dynamic and statistical characteristics of the continuous and bound states of $^2$H$^3$He cluster system.

At the same time, we have successful experience of exploiting the modified potential cluster model (MPCM) based on the including of the Pauli forbidden states (FS) with corresponding classification by Young diagrams [7-16]. This model is much more simple and transparent comparing the microscopic resonating group method (RGM) [17,18]. Nevertheless, MPCM enabled to obtain the consistent numerical results on the majority of the radiative capture reactions total cross sections and rates at astrophysical and thermonuclear energies for more than 30 processes treated in [7-16,19], as well as binding energies and low lying excited spectra, root mean square charge and mass radii, and asymptotic normalizing coefficients (ANC) in cluster channels.

Here we are presenting new calculated data on the root mean square radii and ANC for the $^5$Li in $^2$H$^3$He channel. Total cross section of the deuteron radiative capture on $^3$He to the ground state (GS) of $^5$Li for the dipole $E$1 and $M$1 transitions both in doublet and quartet spin channels have been calculated also. For the obtained astrophysical $S$-factor and reaction rate the analytical parametrizations were found as functions of $E$ and $T_9$ respectively.

---
[i] Corresponding author

## 2. Model and methods

The preliminary analysis of the deuteron radiative capture process on $^3$He in MPCM have been carried out in our early works [20,21], and now near 20 years later new data on the nuclear spectra appeared [22]. In particular, the data on the energy levels of $^5$He and $^5$Li changed drastically comparing the values in [23] those we used in our previous papers.

Modern results are presented in the review [22] and on the website of the authors of this and other similar reviews on TUNL nuclear data [24]. The earlier position of the $P_{1/2}$ level in $^5$Li was pointed as 9 MeV with 5 MeV width at the threshold 16.4 MeV in $^2$H$^3$He channel [23]. However, the new data [22] reported the $P_{1/2}$ energy level at 1.49 MeV and its width equals 6.6 MeV, as well as the threshold value is assumed now as 16.66 MeV. That is why taking into account these new data we obtained another $^2$H$^3$He interaction potentials comparing the previously ones obtained in [20,21]. The additional information on the elastic scattering phase shifts in the energy range 0–5 MeV from [25] was used.

Detailed description of using model MPCP may be found in [7-16]. One of its modification is taking into account the mixing of the cluster states by Young diagrams $\{f\}$. The splitting of the interaction potentials by Young diagrams has been suggested in [26]. Therefore, the constructed interaction potentials both for the discrete and continuous spectra depending on $\{f\}$ lead to the mixing of the corresponding cluster relative motion wave functions (WFs) by orbital diagrams. This procedure is applying below to the treating of the $^2$H$^3$He system. We found early the signature of these special multiplet splitting features in the calculations of the total radiative capture cross sections for the $N^2$H [10,26], $p^3$H [27], and $^2$H$^3$He systems [21,26]

All general relations for the calculation of the different nuclear characteristics within the present model are given in [7,19]. In particular, the following expressions for the total cross sections have been used

$$\sigma_c(NJ, J_f) = \frac{8\pi K e^2}{\hbar^2 q^3} \frac{\mu}{(2S_1+1)(2S_2+1)} \frac{J+1}{J[(2J+1)!!]^2} A_J^2(NJ, K) \cdot \sum_{L_i, J_i} P_J^2(NJ, J_f, J_i) I_J^2(J_f, J_i). \quad (1)$$

Here $\mu$ is reduced mass in initial channel, $q$ is wave number in fm$^{-1}$ related to the c.m. energy as $q^2 = \frac{2\mu E}{\hbar^2}$; $S_1$, $S_2$ particle spins in initial channel; $K$, $J$ – wave number and momentum of emitted $\gamma$ quantum; $N$ denotes electric $E$ or magnetic $M$ transitions of rank $J$ from the initial state $J_i$ to the final one $J_f$.

For electric convection $EJ(L)$ transitions ($S_i = S_f = S$) there are the following expressions for $P_J$, $A_J$ and $I_J$ in (1)

$$P_J^2(EJ, J_f, J_i) = \delta_{S_i S_f}[(2J+1)(2L_i+1)(2J_i+1)(2J_f+1)](L_i 0 J 0 | L_f 0)^2 \begin{Bmatrix} L_i & S & J_i \\ J_f & J & L_f \end{Bmatrix}^2,$$

$$A_J(EJ, K) = K^J \mu^J \left( \frac{Z_1}{m_1^J} + (-1)^J \frac{Z_2}{m_2^J} \right), \qquad I_J(J_f, J_i) = \langle \chi_f | R^J | \chi_i \rangle. \quad (2)$$

Here $S_i, S_f, L_i, L_f, J_i, J_f$ are the corresponding momentums in initial and final states; $m_1$, $m_2$, $Z_1$, $Z_2$ – masses and charges in initial channel; $I_J$ is the overlapping integral over the radial functions for the scattering $\chi_i$ and bound $\chi_f$ states, depending on the cluster-cluster relative coordinate $R$.

For the spin dipole magnetic $M1(S)$ transition ($S_i = S_f = S$, $L_i = L_f = L$) there are the following expressions for $P_J$, $A_J$ and $I_J$ in (1)

$$P_1^2(M1, J_f, J_i) = \delta_{S_i S_f} \delta_{L_i L_f} \left[ S(S+1)(2S+1)(2J_i+1)(2J_f+1) \right] \begin{Bmatrix} S & L & J_i \\ J_f & 1 & S \end{Bmatrix}^2,$$

$$A_1(M1, K) = i \frac{\hbar K}{m_0 c} \sqrt{3} \left( \mu_1 \frac{m_2}{m} - \mu_2 \frac{m_1}{m} \right), \qquad I_J(J_f, J_i) = \left\langle \chi_f \left| R^{J-1} \right| \chi_i \right\rangle. \tag{3}$$

Here $m = m_1 + m_2$, $\mu_1$, $\mu_2$ are the clusters magnetic moments, all other notations are the same as in (2). For the light clusters $\mu(^2\text{H}) = 0.857438$, $\mu(^3\text{H}) = 2.978662$, $\mu(^3\text{He}) = -2.127625$ [5]. The exact values for the particle' masses have been used: $m(^2\text{H}) = 2.014102$ a.m.u., $m(^3\text{H}) = 3.016049$ a.m.u., and $m(^3\text{He}) = 3.016029$ a.m.u. [28]. Constant $\hbar^2 / m_0$ is equal to 41.4686 MeV fm$^2$ where $m_0$ is the atomic mass unit (a.m.u). Coulomb parameter $\eta = \frac{\mu z_1 z_2 e^2}{\hbar^2 q} = 3.44476 \cdot 10^{-2} \frac{\mu z_1 z_2}{q}$. Coulomb potential for the point-like particles is of the form $V_c(\text{MeV}) = 1.439975 \frac{z_1 z_2}{R}$. We are giving here the numerical values as the accuracy of the further calculations depends strongly over them, especially in the resonance energy ranges, as well as the $^5$Li binding energy in $^2$H$^3$He cluster channel. All numerical methods and computing programs may be found in [7,19].

## 3. Elastic $^2$H$^3$He and $^2$H$^3$H scattering and bound states

Let us present the classification by orbital symmetries of $^2$H$^3$He and $^2$H$^3$H systems, i.e. 2+3 nucleons configuration. The doublet channel spin ($S=1/2$) scattering states depend on two allowed orbital Young diagrams {41} and {32}, so they are regarded to be mixed by the orbital symmetries. The quartet channel spin ($S=3/2$) allows only one symmetry {32}, so these states are pure by the Young diagrams.

Contrary, the ground discrete states of $^5$He and $^5$Li nuclei assume to be pure {41} state [26]. That is why the different interaction potentials in the discrete and scattering states should be also different relatively the Young diagram symmetry.

Finally, the dependence not only on $JLS$ quantum numbers, but on $\{f\}$ orbital symmetry also is taken into account for the nuclear interaction potentials of the attractive Gauss form and exponential part modeling the long-range repulsion [7-16]

$$V(r, JLS\{f\}) = V_{0, JLS\{f\}} \exp(-\alpha_{JLS\{f\}} r^2) + V_{1, JLS\{f\}} \exp(-\beta_{JLS\{f\}} r) + V_c(r) \tag{4}$$

with the point-like Coulomb term $V_c$ defined above.

We oriented on the $^2$H$^3$He elastic scattering phase shifts known in the energy range 0–5 MeV [25]. Comparing the early research [20,21] carried for the energies up to 40 MeV, we changed the potential parameters slightly to fit better the low energy region purposely. The results are given in table 1 for $^2$H$^3$He and $^2$H$^3$H systems as the difference is coming in Coulomb interaction only at the same symmetry classification. For the both systems we put repulsive part to be $V_1 = 0$. The last column tells the binding states (BSs) energies. Table 1 shows the binding energies for the quartet forbidden state with the Young diagram {5} and orbital momentum $L = 0$, as well as state with diagram {41} and $L = 1$. The allowed state corresponding the symmetry $\{f\} = \{32\}$ with $L = 0$ and 2 turned to be unbound, i.e. lying in continuous spectrum as corresponds two exciting quanta [26].

Table 1. Potential parameters (4) for the $^2H^3He$ and $^2H^3H$ systems for the mixed by Young diagrams scattering states and corresponding binding energies $E_{BS}$

| $^{2S+1}L$ | $V_0$, MeV | $\alpha$, fm$^{-2}$ | $E_{^2H^3He}$ ($E_{^2H^3H}$), MeV |
|---|---|---|---|
| $^2S_{1/2}$, $^2D$ | -30.0 | 0.15 | -7.0 (-7.9) |
| $^2P$ | -48.0 | 0.1 | -9.6 (-10.2) |
| $^4S_{3/2}$, $^4D$ | -34.5 | 0.1 | -13.0 (-13.9) |
| $^4P$ | -29.0 | 0.1 | -1.4 (-1.8) |

While using the potential $V_0$ = -25.0 MeV, $\alpha$ = 0.15 fm$^{-2}$ from [20,21] for the description of the doublet mixed by the Young diagrams $^2S$ phase shifts one can see in Figure 1 (solid curves for $^2S$ and $^2D$ states) some discrepancy with experimental data [25]. That is why we assumed the potential might be deeper and equals -30.0 MeV (see Table 1), the corresponding results are shown in Figure 1 by dashed curves.

The results for the doublet $^2P$ potential -44.0 MeV, $\alpha$ = 0.1 fm$^{-2}$ from [20,21] show a worse fit of the experimental data in Figure 1 (solid curve) comparing the calculations with deeper potential $V_0$ = -48.0 MeV (dashed curve).

As it is seen from Table 1 in allowed doublet $^2P$ channel the energy of the GS and the first excited state (FES) of $^5Li$ and $^5He$ (experimental spectra is shown in Figure 2) cannot be reproduced with these parameter sets. So, to fit the BSs characteristics of $^5Li$ we found in [20,21] the potential parameters given in Table 2.

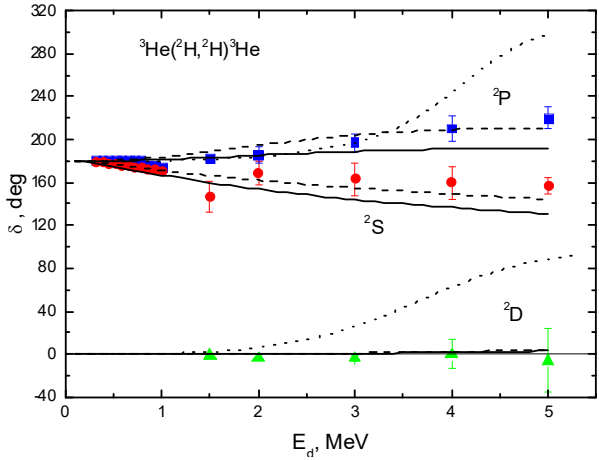
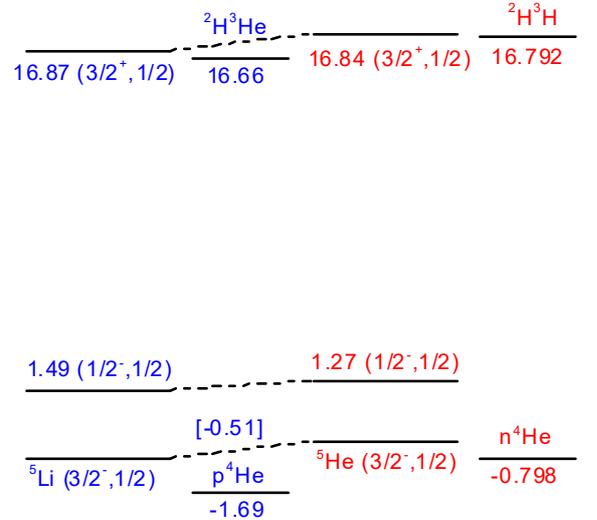

Figure 1. Comparison of the $^2H+^3He$ doublet phase shifts mixed by orbital diagrams calculated with the potentials from Table 1 with results from [25].

Figure 2. Experimental energy spectra for $^5Li$ and $^5He$ with data from [22].

Table 2. Potential parameters for $^2H^3He$ and $^2H^3H$ systems for the pure by Young diagrams states and binding energies $E_{BS}$ from [20,21].

| $^{2S+1}L_J$ | $V_0$, MeV | $V_1$, MeV | $\beta$, fm$^{-1}$ | $E_{^2H^3He}$ ($E_{^2H^3H}$), MeV |
|---|---|---|---|---|
| $S_{1/2}$, $D$ | -40.0 | +8.0 | 0.2 | -8.7 (-9.6) |
| $P_{3/2}$ | -75.5 | – | – | -16.5 (-17.3) |
| $P_{1/2}$ | -60.2 | – | – | -9.0 (-9.7) |

Let us note, that both channel spin states $S$ = 1/2 and 3/2 allow the bound states in $^5He$ or $^5Li$

with total momentum $J = 3/2^-$ (GS) as well as $J = 1/2^-$ (FES) corresponding the P wave. Thus, these both states are the $^{2+4}P$ mixture of singlet and quartet channels. At a time pure doublet states corresponds to {41} diagram, and pure quartet to {32}. Thus, $^{2+4}P$ state may be treated as mixture by Young diagrams also. We see the direct correspondence "channel spin" – "Young diagram", that is why the obtained bellow potentials for the GS and FES with $J = 3/2^-$ and $J = 1/2^-$ we will refer as the pure by Young diagram potentials.

Basing on the latest data for the energy levels [22], as well as refined values for the masses of involved clusters (see Sect. 2) we revised the potential parameters and as it is seen from Table 3 the binding energies in GS and FES of $^5$Li have been reproduced with high accuracy $10^{-6}$ MeV [10]. Note, repulsive potential was taken $V_1 = 0$, and width parameter of $\alpha = 0.18$ fm$^{-2}$ in potential (4). Experimental values for the energy levels are given in brackets next to the calculated $E_{BS}$ [22].

The dimensionless asymptotic normalizing coefficients $C_W$ are given in the last column in Table 3. They are defined according [29]

$$\chi_L(R) = \sqrt{2k_0} C_W W_{-\eta L+1/2}(2k_0 R). \qquad (5)$$

Here $\chi_L(R)$ is numerical GS radial WF, viz. the solution of the Schrodinger equation normalized to unit, $W_{-\eta L+1/2}(2k_0 R)$ is Whittaker function, $k_0$ is wave number related to the channel binding energy. The pointed ANC error is determined by its averaging over interval from 5-6 up to 8-10 fm. The charge radii $r_{rms}$ for the BSs of $^5$Li in the $^2$H$^3$He channel were also calculated and are given in Table 3.

Table 3. New potential parameters for $^2$H$^3$He system for the pure by Young diagrams states

| $L_J$ | $V_0$, MeV | $E_{BS}$ ($E_{exp}$), MeV | $r_{rms}$, fm | $C_W$ |
|---|---|---|---|---|
| $P_{3/2}$ | -84.03570 | -16.660002 (-16.66) | 2.25 | 6.40(1) |
| $P_{1/2}$ | -81.02697 | -15.170001 (-15.17) | 2.26 | 5.83(1) |

We want now attract attention to the first resonance in $^2$H$^3$He state with the width $\Gamma_{cm} = 0.959$ MeV indicated at 19.28 MeV relatively the GS, or at 2.62 MeV relatively the channel threshold and identified as $J^\pi = 3/2^-$ state (see Table 4.3 in [22]). It may be correlated to $P_{3/2}$ wave in doublet or quartet spin channel at 4.37 MeV (l.s.) deuteron energy. It will be shown bellow, that $M1$ transitions from the $P_{3/2}$ scattering states close to this energy are of the resonating character comparing the transitions from the non-resonating $P_{1/2}$ and $P_{5/2}$ waves.

The phase shift analyses carried out in [25] up to the 5 MeV (l.s.) did not reveal any resonating behavior of $^{2+4}P$ waves in spite of rather large width of the considering level. The rising tendency only is seen in $^2P$ phase shift in doublet channel, but quartet $^4P$ phase shift is obviously decreasing. Early we did not treat these resonances, so potentials in Table 1 do not allow to reproduce them. To represent the $^{2+4}P_{3/2}$ resonances the following potential was found

$$V_0 = -1505.3 \text{ MeV and } \alpha = 2.5 \text{ fm}^{-2}. \qquad (6)$$

Calculated $P_{3/2}$ phase shift illustrated by dots in Figure 1 shows the resonance, i.e. equals 90.0(1)°, at 4.37 MeV (l.s.) with the width $\Gamma_{cm} = 1.06$ MeV.

Furthermore, there are two wide resonances with $\Gamma_{cm} = 3.28$ and 4.31 MeV located at 19.45 and 19.71 MeV relatively the GS (or 2.79 and 3.05 MeV relatively the channel threshold) associated to $J^\pi = 7/2^+$ and $J^\pi = 5/2^+$ states (see Table 4.3 in [22]), that may refer to $D$ phase shifts resonating at 4.65 and 5.08 MeV (l.s.) deuteron energies. The second of these resonances in doublet and quartet channels may reveal in $E1$ transitions to the GS. To reproduce such $^{2+4}D_{5/2}$ behavior the following potential was found

$$V_0 = -25.5745 \text{ MeV and } \alpha = 0.075 \text{ fm}^{-2}. \tag{7}$$

Calculated $D_{5/2}$ phase shift illustrated by dots in Figure 1 shows the resonance, i.e. equals 90.0(1)°, at 5.08 MeV (l.s.) with the width $\Gamma_{cm} = 4.43$ MeV. Again there is no such a resonance in $D$ phase shifts analysis done in [25] at mentioned energies, but the tendency of moderate rising of $^4D$ phase shift at 5.0 MeV. We will to conclude, that the revised procedure of processing of the scattering experimental data, obtained in [25], is strongly desirable with extension of the deuteron energy range up to 7–8 MeV and purposely oriented on the check of the signature of the discussed rather wide resonances.

The complete set of transition amplitudes accounted in our calculations is given in Table 4. Transitions from the resonating waves with the main input to the total cross sections are marked as bold. All other transitions from the non-resonating waves (ordinary type) are giving the minor contribution according our estimations. Scattering states corresponding to the same angular momentum $J$, but mixed by channel spin are of the same color.

Table 4. Transitions accounted for the calculated total radiative $^2$H$^3$He capture cross section.

| No. | $^{2S+1}L_J$, initial | Transition | $^{2S+1}L_J$, final |
|---|---|---|---|
| 1. | $^2S_{1/2}$ | $E1$ | $^2P_{3/2}$ |
| **2.** | $^4S_{3/2}$ | **$E1$** | $^4P_{3/2}$ |
| 3. | $^2D_{3/2}$ | $E1$ | $^2P_{3/2}$ |
| **4.** | $^2D_{5/2}$ | **$E1$** | $^2P_{3/2}$ |
| 5. | $^4D_{1/2}$ | $E1$ | $^4P_{3/2}$ |
| 6. | $^4D_{3/2}$ | $E1$ | $^4P_{3/2}$ |
| **7.** | $^4D_{5/2}$ | **$E1$** | $^4P_{3/2}$ |
| 8. | $^2P_{1/2}$ | $M1$ | $^2P_{3/2}$ |
| **9.** | $^2P_{3/2}$ | **$M1$** | $^2P_{3/2}$ |
| 10. | $^4P_{1/2}$ | $M1$ | $^4P_{3/2}$ |
| **11.** | $^4P_{3/2}$ | **$M1$** | $^4P_{3/2}$ |
| 12. | $^4P_{5/2}$ | $M1$ | $^4P_{3/2}$ |

Let us comment the transitions accounted in the calculations of the total cross section of $^3$He($^2$H,$\gamma$)$^5$Li reaction. Since the GS is mixed by spins, dipole $E1$ transition should be treated as occurring from the doublet and quartet $S$ and $D$ scattering states. Within the using model it is impossible to separate the $^2P_{3/2}$ and $^4P_{3/2}$ components explicitly in GS, so we are using the mixed by spin states $P_{3/2}$ function obtained as the solution of Schrodinger equation with potentials from Table 3. For the scattering states the doublet and quartet mixed by Young diagrams calculated with potentials of Table 1 are using.

The interaction potentials have been corroborated by the experimental data on the elastic scattering phase shifts and energy levels spectra, so, WFs obtained as the solutions of the Schrodinger equation with those potentials account effectively the cluster system states, in particular, the mixing by channel spin. Therefore, the total cross section of $E1$ transition from the mixed continuous states in to also spin mixed GS may be taken as simple doubling of the partial cross section as each is calculated with the same functions, however, spin algebraic factors are specified for every matrix element [7-16]. In reality, there is only one transition from the scattering state to the GS, rather than two different $E1$ processes.

The averaging procedure concerns the transitions from $D_{5/2}$ and $D_{3/2}$ scattering states to the $P_{3/2}$ GS of $^5$Li in $^2$H$^3$He channel. Finally, we arriving to the following $E1$ multipole cross section

$$\sigma_0(E1) = \sigma(^2S_{1/2} \to {}^2P_{3/2}) + \sigma(^4S_{3/2} \to {}^4P_{3/2}) + \sigma(^4D_{1/2} \to {}^4P_{3/2}) +$$
$$+ [\sigma(^2D_{3/2} \to {}^2P_{3/2}) + \sigma(^4D_{3/2} \to {}^4P_{3/2})]/2 + [\sigma(^2D_{5/2} \to {}^2P_{3/2}) + \sigma(^4D_{5/2} \to {}^4P_{3/2})]/2.$$

According the classification in Table 4 there are also spin mixing states in scattering $P$ waves leading to the magnetic dipole $M1$ transition, so the corresponding cross section is written as above one for the $E1$ transition to the GS

$$\sigma_0(M1) = \sigma(^4P_{5/2} \to {}^4P_{3/2}) + [\sigma(^2P_{1/2} \to {}^2P_{3/2}) + \sigma(^4P_{1/2} \to {}^4P_{3/2})]/2 +$$
$$+ [\sigma(^2P_{3/2} \to {}^2P_{3/2}) + \sigma(^4P_{3/2} \to {}^4P_{3/2})]/2.$$

Thus, we have identified all the major transitions that may contribute to the total cross sections of the deuteron capture process on $^3$He at low energies, which are treated in this paper.

### 4. Total cross section, astrophysical $S$-factor and reaction rate

Figure 3 shows the results of the calculated $E1$ radiative capture in $^2$H$^3$He cluster channel at the energies bellow 5 MeV. Solid red line denotes the cross section for the $E1$ transition from the $^2S$ and $^4S$ scattering waves (potentials from Table 1) to the GS $^{2+4}P_{3/2}$ defined by the interaction potential parameters from Table 3. Cross sections for the $E1$ transitions from the $^2S$ wave is of few orders suppressed as it is of non-resonant behavior.

Solid violate line in Figure 3 denotes the cross section for the $E1$ transition to the GS from the resonating $^{2+4}D_{5/2}$ waves calculated with potential (7) and includes all other small in value amplitudes for the non-resonating $D$ waves listed in Table 4.

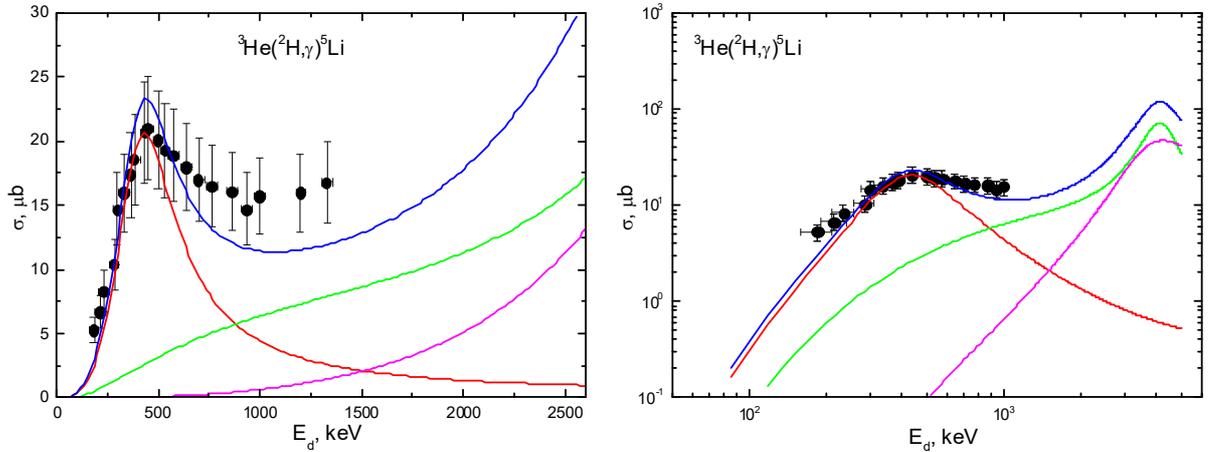

Figure 3a. Total cross section for $^3$He($^2$H,γ)$^5$Li bellow 2.5 MeV. Experimental data from [3]. Calculated curves correspond to the potential parameters of Tables 1 and 3.

Figure 3b. Total cross section for $^3$He($^2$H,γ)$^5$Li bellow 5.0 MeV. Data are the same as in Figure 3a.

Green curve in Figure 3 shows the contribution of $M1$ transitions from the resonating $^{2+4}P_{3/2}$ waves corresponding to the potential (6) and non-resonating set for $P$ potentials from Table 1. Note, that M1 transitions from non-resonant scattering P waves have a significant impact on the total cross sections only at the energies above 600-700 keV

The total cross section included all transitions listed in Table 4 is shown by blue curve in Figure 3. It is well seen, that starting from the energies 600-800 keV the calculated cross section is lying a little bit lower the error bars band. Note, those errors have been taken equal to 19%, while we oriented on the value $\sigma = 21(4) \mu b$ at 450 keV within the 25 keV energy scaling error pointed in [3].

Figure 4 displays the calculated astrophysical *S*-factor in direct correspondence with the cross sections shown in Figure 3. We recalculated data on the cross sections from [3] into *S*-factor and presented them as points in this figure. As we defined at minimal energies 185–300 keV its value is near 0.39(5) keV·b. This value may be approximated by trivial constant energy dependence $S(E) = S_0$ with $S_0 = 0.386$ keV·b and mean value for $\chi^2 = 0.21$. Same experimental 19% errors were assumed for *S*-factor. So, the linear parametrization at the energies bellow 20 keV is shown by dashed green line in Figure 4.

To improve the description of the experimental data we tried the following approximating function

$$S(E) = S_0 + S_1 E + S_2 E^2, \qquad (8)$$

but did not succeed at this very low energy region. So, the obtained *S*-factor turned to be rather stable in the energy range 20–50 keV and equals to 0.14(1) keV·b, that is essentially less than experimental values in [3]. Pointed here error for the calculated *S*-factor was defined as its averaging over energy interval 20–50 keV.

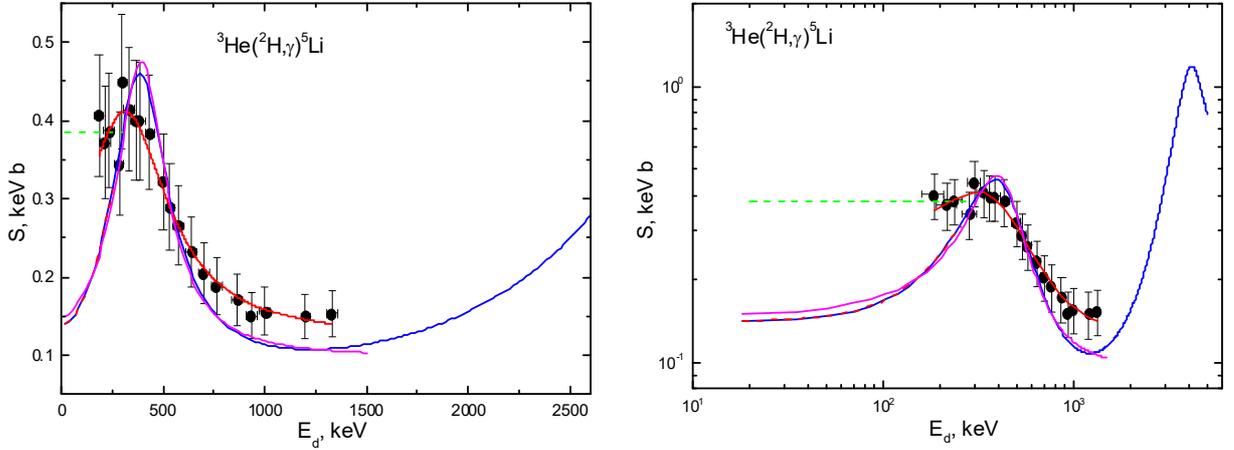

Figure 4a. *S*-factor data from [3] and fits with potentials from Tabs. 1 and 3 for $^3\text{He}(^2\text{H},\gamma)^5\text{Li}$ bellow 2.5 MeV.

Figure 4b. Same as Figure 4a, but at the energy bellow 5.0 MeV.

In what follows we implemented the parametrization of the calculated *S*-factor according the expression (8) with $S_0 = 0.14081$ keV·b, $S_1 = -2.0705 \cdot 10^{-5}$ b, $S_2 = 2.7897 \cdot 10^{-6}$ b·keV$^{-1}$. We found the value $\chi^2$ to be 0.17 within 1% precision of the theoretical *S*-factor. The result is shown by red dashed curve in Figure 4 and is consistent with experimental data in the energy region close to 250 keV.

We did apply the ordinary $\chi^2$ statistics as usually was done in [7,19] and defined as

$$\chi^2 = \frac{1}{N}\sum_{i=1}^{N}\left[\frac{S^a(E_i) - S^c(E_i)}{\Delta S^c(E_i)}\right]^2 = \frac{1}{N}\sum_{i=1}^{N}\chi_i^2 .$$

Here $S^c(E_i)$ are the data for the calculated astrophysical *S*–factor corresponding the blue curve in Figure 4. Data for the approximating value $S^a(E_i)$ are taken according formula (8). Then, the error $\Delta S^c(E_i)$ is assumed to be 1%, *N* is the number of taken into account points.

Experimental data shown by dots in Figure 4 may be approximated by the function of Breit-Wigner type

$$S(E) = a_1 + \frac{a_2}{(E-a_3)^2 + a_4^2/4}$$

with the following parameters $a_1 = 0.12315$, $a_2 = 18861$, $a_3 = 313.22$, $a_4 = 509.95$. The results of this parametrization is shown by solid red curve in Figure 4, $\chi^2$ is equal to 0.11.

This very approximation form was used in the energy interval up to 1.5 MeV, but with the parameter set 0.096846, $a_2 = 8520.4$, $a_3 = 393.13$, $a_4 = 300.34$. The quality of the fit with $\chi^2 = 14.5$ and errors within 1% is illustrated by the violet solid curve in Figure 4. Note, the value of the approximated $S$-factor at 20 keV is equal to 0.15 keV·b.

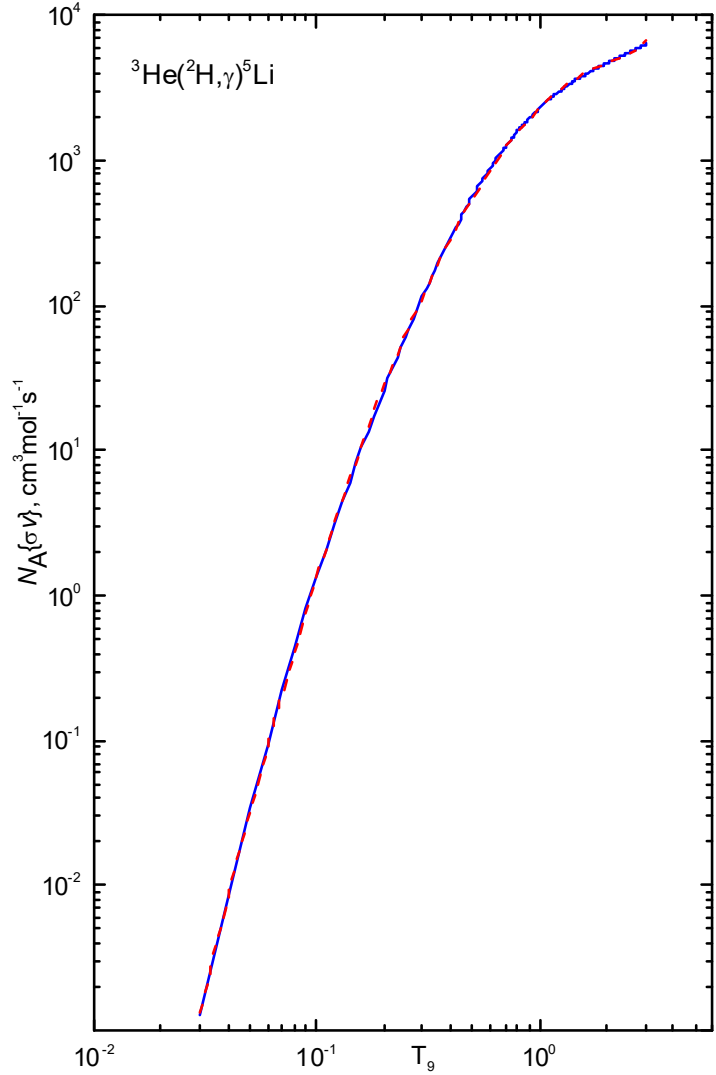

Figure 5. Reaction rate of the deuteron radiative capture on $^3$He. Theoretical curves were obtained with the potentials from Tabs. 1 and 3. An explanation of the parameterization functions is given in Sect. 4.

Figure 5 shows the calculated reaction rate for the deuterium radiative capture on $^3$He for the temperatures from 0.03 up to 3 $T_9$. The blue curve was obtained basing on the corresponding theoretical cross section given in Figure 3. The last one differs slightly from the experimental data, but we found no measured cross sections at higher energies at least up to 5 MeV [4], or may be any calculated rates by other authors.

The reaction rate in cm$^3$mol$^{-1}$s$^{-1}$ units may be presented as usual [30]

$$N_A \langle \sigma v \rangle = 3.7313 \cdot 10^4 \mu^{-1/2} T_9^{-3/2} \int_0^\infty \sigma(E) E \exp(-11.605 E / T_9) dE,$$

where the energy $E$ is taken in MeV, total cross section $\sigma(E)$ in $\mu b$, reduced mass $\mu$ in a.m.u., and temperature $T_9$ in $10^9$ K. To calculate this integral 2000 points of the theoretical cross section were taken in the c.m. energy range from 1 to 2000 keV. Expansion of this interval to 3 MeV and number of points to 3000 is changing of the reaction rate value less than 1% order.

The calculated reaction rate was approximated in the range $0.03 - 3.0$ $T_9$ as a following

$$N_A \langle \sigma v \rangle = 8.2838 \cdot 10^5 / T_9^{2/3} \cdot \exp(-6.6685 / T_9^{1/3}) \cdot (1.0 - 3.6550 \cdot T_9^{1/3} - 1.2850 \cdot T_9^{2/3} + 28.989 \cdot T_9 - 33.646 \cdot T_9^{4/3} + 10.779 \cdot T_9^{5/3}). \tag{9}$$

The resulting curve is shown as red one in Figure 5, $\chi^2$ is equal to 7.2. To find the parameters in (9) we used 300 points corresponding the blue curve in the same figure. To estimate the $\chi^2$ the error was taken to be 1%.

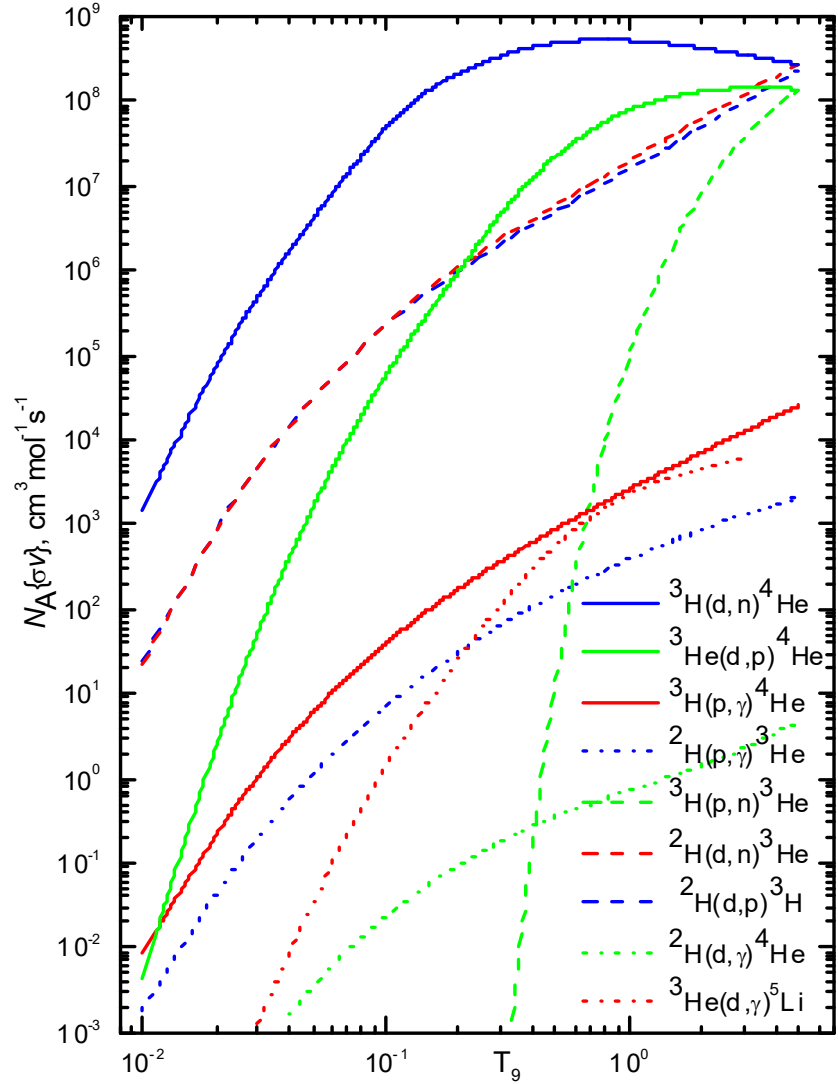

Figure 6. Comparison the reaction rates, approximated in [31] and obtained here.

We see the main goal of present research to determine the role of radiative capture reaction $^3$He($^2$H,$\gamma$)$^5$Li in the balance of the processes with the deuterons occurring in the laboratory and nature plasma. We based on the data [31] for the parametrization of reaction rates involving the lightest and light nuclei. Present contribution in to this compilation, important from the practical application ansatz is the calculated reaction rate for the process $^3$He($^2$H,$\gamma$)$^5$Li and its analytical parametrization (9). Figure 6 displays the comparative rates of the deuteron and proton induced reactions calculated according the parametrizations from [31], as well as present results. It is obvious the treated reaction is near four orders less in magnitude comparing the $^3$H(d,n)$^4$He process, that is giving the essential contribution to the primordial nucleosynthesis of lightest elements in the Universe together with the reactions $^3$He(d,p)$^4$He, $^2$H(d,n)$^3$He, as well as $^2$H(d,p)$^3$H.

## 5. Conclusion

Comparatively simple model representations succeeded to obtain the theoretical results in general agreement with the available experimental data for the $S$-factor, except first three points at 185–235 keV, and at the energies higher 600–800 keV.

The minimum value 14.5(2.8) $\mu b$ of the experimental cross section is close to 1 MeV [3], and the calculated one is 11.5 $\mu b$. We see no reasonable explanation of this discrepancy between theory and experiment yet. Despite 12 transitions listed in Table 4 have been taken into account, we do not have considered some kind other features of this process.

For example, the small admixture of $^4F_{3/2}$ component comparing the dominating $^4P_{3/2}$ in GS of $^5$Li may give any input into the total cross section. We suppose it may increase the total cross section value, specially at higher energies.

Another option to improve the agreement for the obtained results with experiment is including of small admixture of $\alpha N$ component in $d\tau$ cluster channel wave function of $^5$Li, which may affect the asymptotics of the radial function, and, as a consequence, should redistribute the space probability density in the nuclear interior. Same superposition of various cluster components RGM is using [17,18].

The result may be important in practical terms concerns the obtained approximation of the experimental $S$-factor as energy constant 0.39 keV·b with mean $\chi^2$ = 0.21 at 19% experimental error bars below 300 keV. Another approximation by square form (8) was found at the energies bellow 250 keV, it shows $\chi^2$ = 0.17 at 1% calculation errors. Within the energy range from 185 keV up to 1.4 MeV fit by Breit-Wigner resonance-type formula led to $\chi^2$ = 0.11 at 19% experimental error bars. Same formula applied for the approximation of the theoretical curve in the energy interval from 20 keV up to 1.5 MeV gave $\chi^2$ = 14.5 at 1% calculation errors.

For the reaction rate $\sigma v$ we suggested the parametrization (9) for the temperatures from 0.03 up to 3 $T_9$, and would like to recommend it for the solution of some applied problems both in astrophysical and fusion current study.

Finally, it should be noted that we are aware of only one measurement of the total cross sections of this capture reaction presented in [3], which were performed in the late 60-ies of the last century, that is, about 50 years ago, and which may be subject to change. Therefore, for the final conclusions need to have more precise measurements of this reaction, using the modern methods and obtained irrespective previous data. We hope presented results may be a guide for the future experimental proposals.

## Acknowledgments

The work was performed under grant No. 0073/PCF-15-MES "The astrophysical study of stellar and planetary systems" and "Studying of the thermonuclear processes in the Universe" of